\documentclass[a4paper,onecolumn]{article}
\usepackage{amsmath}
\usepackage{graphicx}
\usepackage{braket}
\usepackage{caption}
\usepackage{subcaption}
\usepackage{authblk}
\usepackage{verbatim}

\renewcommand\Re{\operatorname{Re}}
\renewcommand\Im{\operatorname{Im}}

\title{Resummation of not summable series}

\author[1,2]{Zbigniew Ambrozi\'nski\thanks{zbigniew.ambrozinski@uj.edu.pl}}
\author[1]{Jacek Wosiek\thanks{wosiek@th.if.uj.edu.pl}}
\affil[1]{Marian Smoluchowski Institute of Physics, Jagiellonian University\\Reymonta 4, 30-059 Cracow, Poland}
\affil[2]{Max-Planck-Institut f\"ur Gravitationsphysik (Albert-Einstein-Institut)\\Am M\"uhlenberg 1, D-14476 Golm, Germany}

\begin{document}
\maketitle
\abstract{
In this paper we present results of computations for the ground energy of weakly coupled double well potential in quantum mechanics. We give a numerical evidence for cancelation of imaginary contributions to energy coming from Borel resummation and multi-instanton terms. We also estimate several higher coefficients of the multi-instanton expansion which are not given in the literature.}

\section{Introduction}
The question of the relation between the perturbative and
non-perturbative contributions in quantum theories is a long standing one.
The answer emerged over years of studies and is well known, e.g. for the
anharmonic oscillator \cite{ZJ1,Bog}. The two contributions
are in principle additive, however there is a sublte interplay between
them. Namely, the ambiguities in resummation of perturbative, asymptotic
series are non-perturbative and cancel against ones of multi-instanton
contributions. In this paper we verify this claim by confronting
theoretical predictions with, very high precision, numerical solutions of
the problem.

We consider a double well potential in the following parametrization:
\begin{align}
V(x)=\frac{1}{2}x^2(1-\sqrt{g}x)^2.
\end{align}
As it is well known \cite{BGZJ1977}, the perturbation series for the ground energy $E(g)$ is not Borel summable. Still, one can perform a Borel sum for complex coupling constant $g$ and analytically continue it to positive axis from lower or upper half of the complex plane. This freedom results in an ambiguity of imaginary part of the Borel sum. However, there is another ambiguity originating from instanton contributions to energy. As stated in \cite{ZJ1} these imaginary terms must cancel.

In this paper we give a direct numerical evidence of this fact. To this end, we find perturbative series of the ground state energy and construct its Borel transform. Then, we continue it analytically using Pad\'e approximation and perform inverse Borel transform. It is done in the limit $g=\Re g+ i0^+$. Next, we demonstrate that imaginary part of the two--instanton molecule contribution derived by Bogomolny \cite{Bog} cancels imaginary part of energy in Borel sum at leading order. We also confirm cancelation of higher order imaginary terms given by Jentschura and Zinn--Justin in \cite{ZJJ} and find next few coefficients of their expansion.

To compute energies numerically we use, essentially exact, cut Fock space approach \cite{Wos,CW,WT,Korcyl}. With this method we show that real part of the Borel sum plus two--instanton molecule contribution give very accurate approximation to the energy for small couplings.

\section{Borel resummation}
One can find perturbation series of the ground state energy $E(g)=\sum_{k=0}^{\infty}\epsilon_k g^k$ up to high orders using the Rayleigh-Schr\"odinger perturbation theory \cite{BW1}. We found $\epsilon_k$ for $k\leq500$. Asymptotic behavior of $\epsilon_k$ is known \cite{BPZJ} and yields
\begin{align}
\epsilon_k\approx -k! 3^k\frac{3}{\pi}.
\end{align}
The relative difference between asymptotic estimate and exact value of $\epsilon_k$ is $0.6\%$ for $k=500$ and decreases at rate estimated in \cite{BPZJ}. The perturbative series is asymptotic and a resummation procedure is needed. To this end we use the Borel transform
\begin{align}
B_K(t)=\sum_{k=0}^{K} \frac{\epsilon_k}{k!}t^k.
\end{align}
$B_\infty(t)$ is convergent for $|t|<\frac{1}{3}$ and has a pole at $t=\frac{1}{3}$. Both follow from the asymptotic behavior of $\epsilon_k$. Inverse Borel transform is given by the integral
\begin{align}\label{eq:Borel_integral}
E_{Borel}(g)=\frac{1}{g}\int_0^\infty dt e^{-t/g}B(t)
\end{align}
where $B(t)$ is analytic continuation of the Borel transform $B_\infty(t)$. Because the pole at $t=\frac{1}{3}$ lies on the integration path, the perturbative series is called not Borel summable \cite{BGZJ1977}. Still, the integral can be calculated for $t=\Re t\pm i0^+$. One can show that it is equivalent to take $t$ real but $g=\Re g\pm i0^+$. Changing between positive and negative imaginary part of $g$ alters sign of imaginary part of the integral (\ref{eq:Borel_integral}). In particular, contribution of the leading singularity at $t=\frac{1}{3}$ is
\begin{align}\label{eq:Borel_ambiguity}
E_{Borel}(g+i0^+)-E_{Borel}(g-i0^+)=-\frac{2i}{g}e^{-1/3g}.
\end{align}

\begin{figure}
        \begin{subfigure}[b]{0.3\textwidth}
                \centering
                \includegraphics[width=\textwidth]{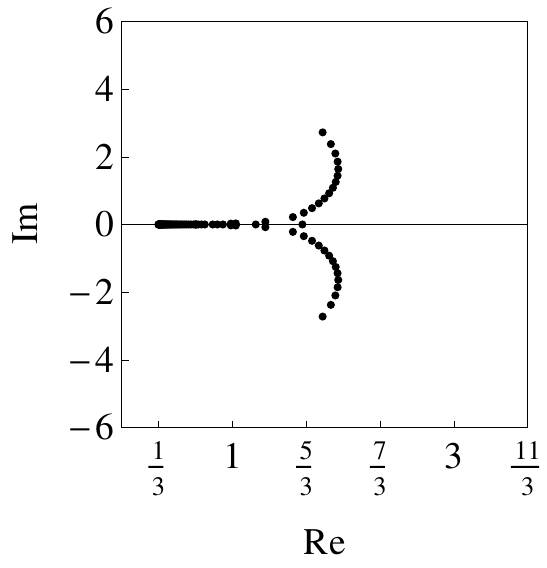}
                \caption{$K=200$}
        \end{subfigure}%
        \begin{subfigure}[b]{0.3\textwidth}
                \centering
                \includegraphics[width=\textwidth]{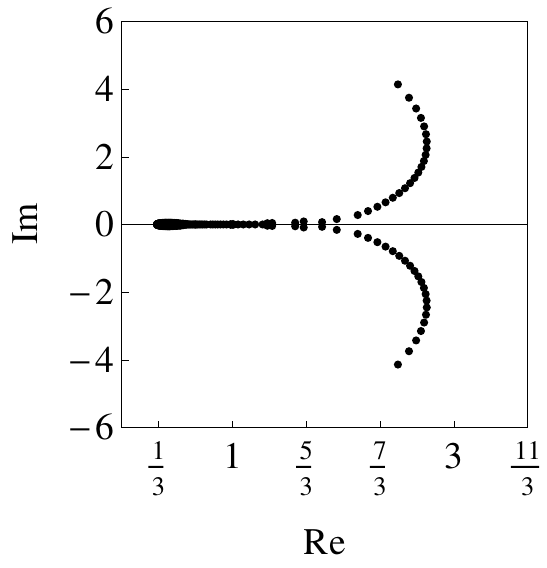}
                \caption{$K=350$}
        \end{subfigure}
        \begin{subfigure}[b]{0.3\textwidth}
                \centering
                \includegraphics[width=\textwidth]{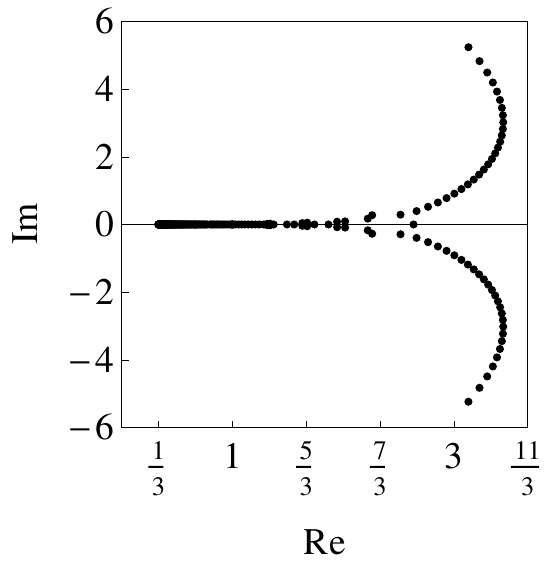}
                \caption{$K=500$}
        \end{subfigure}
        \caption{Poles of Pad\'e approximant for different orders of approximation $K$.}\label{fig:padepoles}
\end{figure}

As an approximation of the analytic continuation $B(t)$ we took the diagonal $\left[\frac{K}{2}/\frac{K}{2}\right]$ Pad\'e approximant $P(t)$ of $B_K(t)$. We shall now analyze poles of $P(t)$. They are presented in Fig. \ref{fig:padepoles}. Poles on the real axis condense with growing $K$ and form a cut on the interval $[\frac{1}{3},\infty)$. They reflect a cut of $B(t)$. Poles with nonzero imaginary part move to infinity as $K$ increases. We infer that they lie in the region where the Pad\'e approximant is no longer reliable. It is known \cite{DG} that the approximant is weakly convergent near poles of the approximated function. Therefore, we changed the integration contour from $t=\Re t+i0^+$ to $t=|t|e^{i\pi/4}$. Secondly, we cut the integral at $\Re t=1.4$ so that we did not come close to poles of $P(t)$. Error coming from both, change of integration contour and cutting the integral is of order $\frac{1}{g}e^{-1.4/g}$. It is much smaller than the ambiguity of Borel sum already for $g=0.1$ and therefore we shall neglect it. Major error for small coupling constant $g$ is an effect of finite $K$. It is a nontrivial task to estimate it a priori. One has to try different $K$'s and check if one can reach such values that the energy is independent of $K$.

\section{Instanton contributions}
The perturbation theory yields the same series expansion of the ground energy in both minima of the potential $V(x)$, $x=0$ and $x=1/\sqrt g$. Thus, the ground energy is degenerate at the level of perturbation theory.

One can calculate splitting of the ground energy $E_0$ and the first excited energy $E_1$ using semiclassical methods in the dilute instanton gas approximation. The difference is $E^{(1)}_1-E^{(1)}_0=\frac{1}{\sqrt{\pi g}}e^{-1/6g}$. The leading correction to dilute gas approximation due to interactions between instantons was later calculated by Bogomolny \cite{Bog}:
\begin{align}
E_N^{(2)}(g)=\frac{1}{\pi g}e^{-1/3g}\left(\gamma+\ln\left(-\frac{2}{g}\right)\right),&&N=0,1,
\end{align}
where $\gamma\approx 0.577$ is the Euler constant.
One has to understand this formula as a continuation from negative $g$ through upper or lower half of the complex plane, i.e. taking $g+i0^+$ or $g-i0^+$ limit. These limits give different results:
\begin{align}\label{eq:molecule_ambiguity}
E_N^{(2)}(g+i0^+)-E_N^{(2)}(g-i0^+)=\frac{2i}{g}e^{-1/3g},&&N=0,1.
\end{align}
Note that this is exactly opposite to the leading order of ambiguity of Borel sum (\ref{eq:Borel_ambiguity}).

Full formula for instanton and perturbative contributions to energy was given by Zinn-Justin \cite{ZJ1,ZJJ}:
\begin{align}\label{eq:full_expansion}
E_N(g)=\sum_{k=0}^\infty \epsilon_k g^k+\sum_{n=1}^\infty\left(-(-1)^N \frac{e^{-1/6g}}{\sqrt{\pi g}}\right)^n\sum_{l=0}^{n-1}\left(\ln\left(-\frac{2}{g}\right)\right)^{l}\sum_{k=0}^\infty \epsilon_{nlk} g^k
\end{align}
for $N=0,1$. Some coefficients were given in \cite{ZJJ}:
\begin{align}
\epsilon_{200}&=\gamma,&\epsilon_{210}&=1,\nonumber\\
\epsilon_{201}&=-\frac{23}{2}-\frac{53}{6}\gamma,&\epsilon_{211}&=-\frac{53}{6},\label{eq:coefficients}\\
\epsilon_{202}&=\frac{13}{2}-\frac{1277}{72}\gamma,&\epsilon_{212}&=-\frac{1277}{72}.\nonumber
\end{align}
Cancelation of ambiguities (\ref{eq:Borel_ambiguity}) and (\ref{eq:molecule_ambiguity}) renders formula (\ref{eq:full_expansion}) unique at least at order $g^{-1}e^{-1/3g}$. We will use numerical analysis to see that the series is unique also at higher orders in $g$, i.e. $n=2$ and $k>0$. We will also show that real part of energy improves when one adds two--instanton terms to the Borel energy (\ref{eq:Borel_integral}). This can be done only if we eliminate much larger contributions from independent instanton $(n=1)$. Note that $n=1$ terms are exactly opposite for the ground and first excited energy. Therefore, we will be interested in their mean values $E=\frac{1}{2}(E_0+E_1)$

\section{Cut Fock space method}
\begin{figure}
\includegraphics[width=\textwidth]{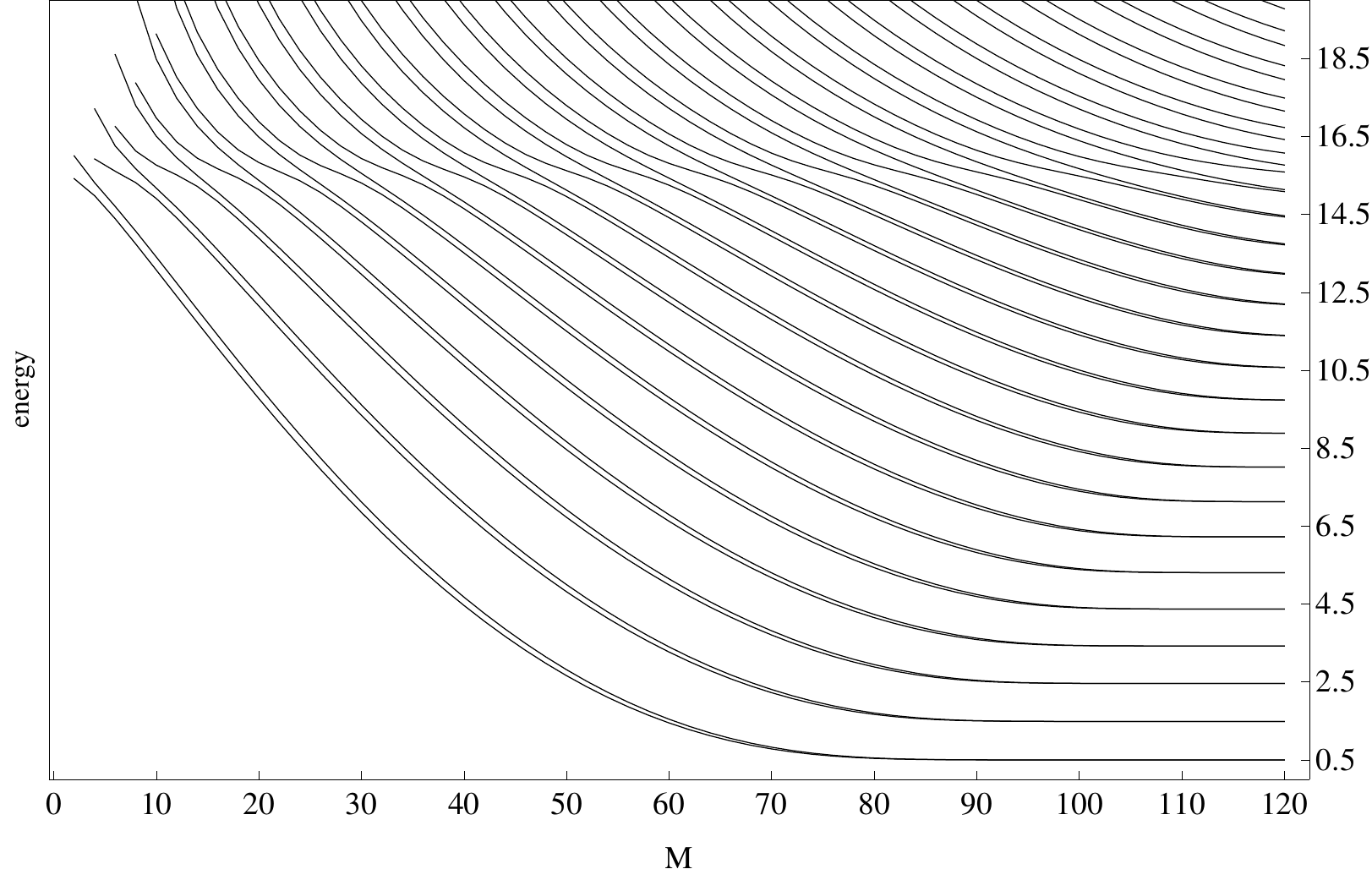}
\caption{Convergence of eigenvalues of the matrix $H_M$ with growing $M$ for $g=0.002$. Behavior of eigenenergies changes
qualitatively around the top of the barrier.}\label{fig:convergence}
\end{figure}

An alternative technique of computing the lowest energies is the cut Fock space method. Let us denote by $\ket{n}$ the Fock basis, which is set of energy states of the harmonic oscillator with minimum at $x=\frac{1}{2\sqrt g}$. Then the matrix $H_M=(\braket{n|\frac{1}{2}P^2+V(X)|m})_{nm}$ with $n,m<M$ is an approximation of the Hamiltonian. $M$ is called the cut--off. We expect that the lowest eigenvalues of $H_M$ approximate energies of the system. It was very well confirmed in many cases \cite{Wos,CW,WT}. For our system convergence of energies with increasing $M$ is presented in Fig. \ref{fig:convergence}.

For small coupling constant $g$ there are several eigenenergies smaller than height of the barrier, e.g. there are 38 for $g=0.002$ with $V(1/2\sqrt g)\approx 15.6$. Classically the potential barrier is impenetrable for states with such energy. In the classical limit, there is a pair of degenerate states, one localized in left and one in right minimum of the potential. Due to quantum tunneling these states mix into symmetric and antisymmetric combinations. The symmetric state of each pair has slightly lower energy than the antisymmetric one.

Let us now analyze convergence of the energies with growing cut--off. Characteristic feature of this method is, as in any variational method, that energies are approximated from above. Therefore, they fall down as the cut--off increases and approximation improves. For small cut--offs energies are higher than the potential barrier and splitting between them is large. Only when energies become smaller than the potential barrier they join into pairs. This can be clearly seen in Fig. \ref{fig:convergence}.

Another feature is that energies fall linearly for small cut--offs. This is because the space basis does not yet explore minima of the potential. For cut--offs high enough the convergence becomes approximately exponential. One can find a more detailed analysis of this system in \cite{Amb}.

The cut Fock space method yields no approximations apart from precision of computations and finite cut--off effects. Thus, it is essentially exact. For our purposes we consider the value
\begin{align}
E_{Fock}(g)=\frac{1}{2}(E_0+E_1)
\end{align}
where $E_{0,1}$ are the two lowest eigenvalues of $H_M$. For the smallest considered value of coupling constant $g=0.00016$ the needed cut--off was $M=5000$ and precision $10^{-{980}}$. The cut Fock space method turns out to be far more efficient than computing energies through Borel resummation procedure. Still higher cutoff and greater numerical precision can be applied resulting in more accurate results.

\section{Comparison}
\begin{figure}
\centering
\includegraphics[width=\textwidth]{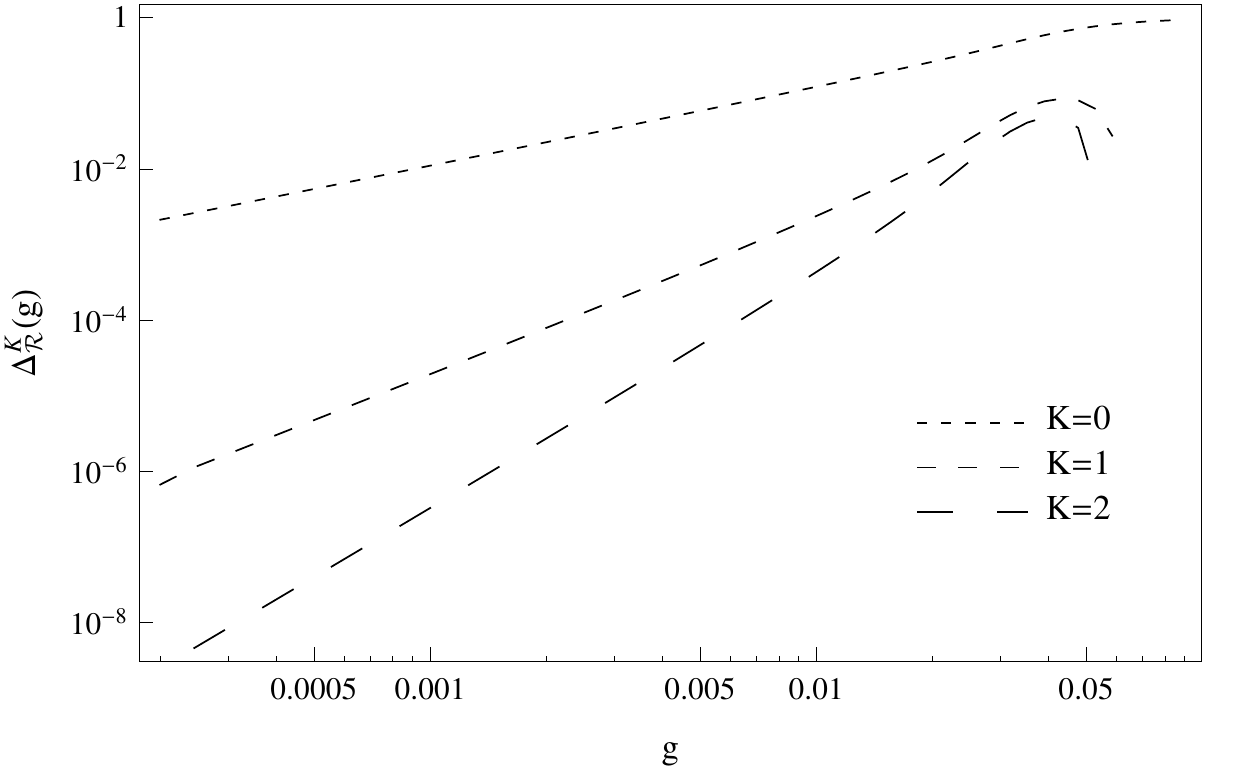}
\caption{Function $\Delta_{\mathcal R}^K(g)$ for $K=0,1,2$. One expects that $\Delta_{\mathcal R}^K(g)=\mathcal O(g^{\alpha_K})$ with $\alpha_K=K+1$. Fitting a straight line to the log-log plot gives $\alpha_{K=0}=1.0242\pm0.0007$, $\alpha_{K=1}=2.05\pm0.02$, $\alpha_{K=2}=3.039\pm0.002$. Fitted values of $\alpha_K$ are slightly above expectance because of higher order corrections, which are not negligible for nonzero $g$.}\label{fig:realconvergence}
\end{figure}

\begin{figure}
\centering
\includegraphics[width=\textwidth]{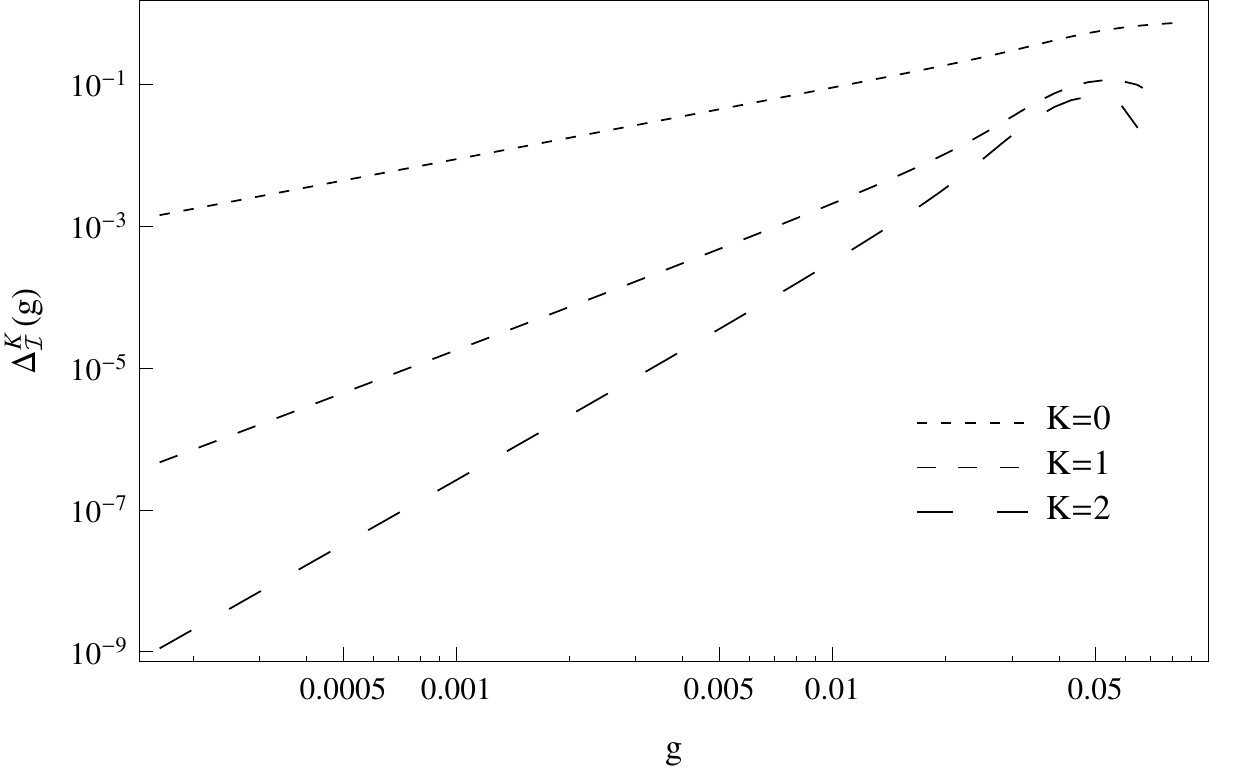}
\caption{Function $\Delta_{\mathcal I}^K(g)$ for $K=0,1,2$. One expects that $\Delta_{\mathcal I}^K(g)=\mathcal O(g^{\alpha_K})$ with $\alpha_K=K+1$. Fitting a straight line to the log-log plot gives $\alpha_{K=0}=1.001\pm0.0001$, $\alpha_{K=1}=2.0075\pm0.0008$, $\alpha_{K=2}=3.009\pm0.0009$.}\label{fig:imaginaryconvergence}
\end{figure}

We will show that $\mathrm{Im} E_{Borel}(g)+\mathrm{Im} E^{(2)}(g)=0$ and $\mathrm{Re} E_{Borel}(g)+\mathrm{Re} E^{(2)}(g)=E_{Fock}(g)$ up to leading order disregarded in $E^{(2)}(g)$. To this end we introduce
\begin{align}
E^{(2),K}&=\frac{1}{\pi g}e^{-1/3g}\sum_{l=0}^1\left(\ln\left(-\frac{2}{g}\right)\right)^l\sum_{k=0}^K\epsilon_{2lk}g^k,\\
\Delta_{\mathcal I}^K(g)&=\frac{\Im E_{Borel}(g)+\Im E^{(2),K}(g)}{\frac{1}{g}e^{-1/3g}},\\
\Delta_{\mathcal R}^K(g)&=\frac{\Re E_{Borel}(g)+\Re E^{(2),K}(g)-E_{Fock}(g)}{\frac{1}{\pi g}e^{-1/3g}\ln(2/g)}.
\end{align}
Denominators of $\Delta_{\mathcal I}^K(g)$ and $\Delta_{\mathcal R}^K(g)$ are leading terms of imaginary and real part of $E^{(2)}(g)$ respectively. According to formula (\ref{eq:full_expansion}) we expect that $\Delta_{\mathcal R,\mathcal I}^K(g)=\mathcal O(g^{K+1})$. Plots of $\Delta_{\mathcal R,\mathcal I}^K(g)$ for $K=0,1,2$ are presented in Figs \ref{fig:realconvergence},\ref{fig:imaginaryconvergence}. Asymptotic behavior of $\Delta_{\mathcal R,\mathcal I}^K(g)$ agrees with predictions. We also found approximations of coefficients (\ref{eq:coefficients}) from numerical data and confirmed coefficients $\epsilon_{21k}$ in with precision $10^{-20}$ and coefficients $\epsilon_{20k}$ with precision $10^{-8}$.

It turned out that it is possible to determine several next coefficients $\epsilon_{2lk}$ in expansion (\ref{eq:full_expansion}). Taking coefficients (\ref{eq:coefficients}) as given we found
\begin{align*}
\epsilon_{203}&=-\frac{45941}{144}-\frac{336437}{1296}\gamma\pm1.6\cdot 10^{-10},\\
\epsilon_{204}&=-\frac{20772221}{2592}-\frac{141158555}{31104}\gamma\pm2\cdot 10^{-6},\\
\epsilon_{205}&=-205496.5847-\frac{17542610737}{186624}\gamma\pm2\cdot 10^{-3},\\
\epsilon_{206}&=6936980.4\pm4.8\\
\epsilon_{213}&=-\frac{336437}{1296}\pm1.3\cdot 10^{-21},\\
\epsilon_{214}&=-\frac{141158555}{31104}\pm4.2\cdot10^{-17},\\
\epsilon_{215}&=-\frac{17542610737}{186624}\pm5.9\cdot10^{-13},\\
\epsilon_{216}&=-2221191.7314262645\pm4.8\cdot10^{-9},\\
\epsilon_{217}&=-58524267.633067\pm2.5\cdot10^{-5},\\
\epsilon_{218}&=-1695080020.213\pm9.5\cdot10^{-2},\\
\epsilon_{219}&=-53461315700\pm1.6\cdot10^{3},\\
\epsilon_{21;10}&=-1823771270000\pm4.8\cdot10^{5}.
\end{align*}

\begin{figure}
\centering
\includegraphics[width=\textwidth]{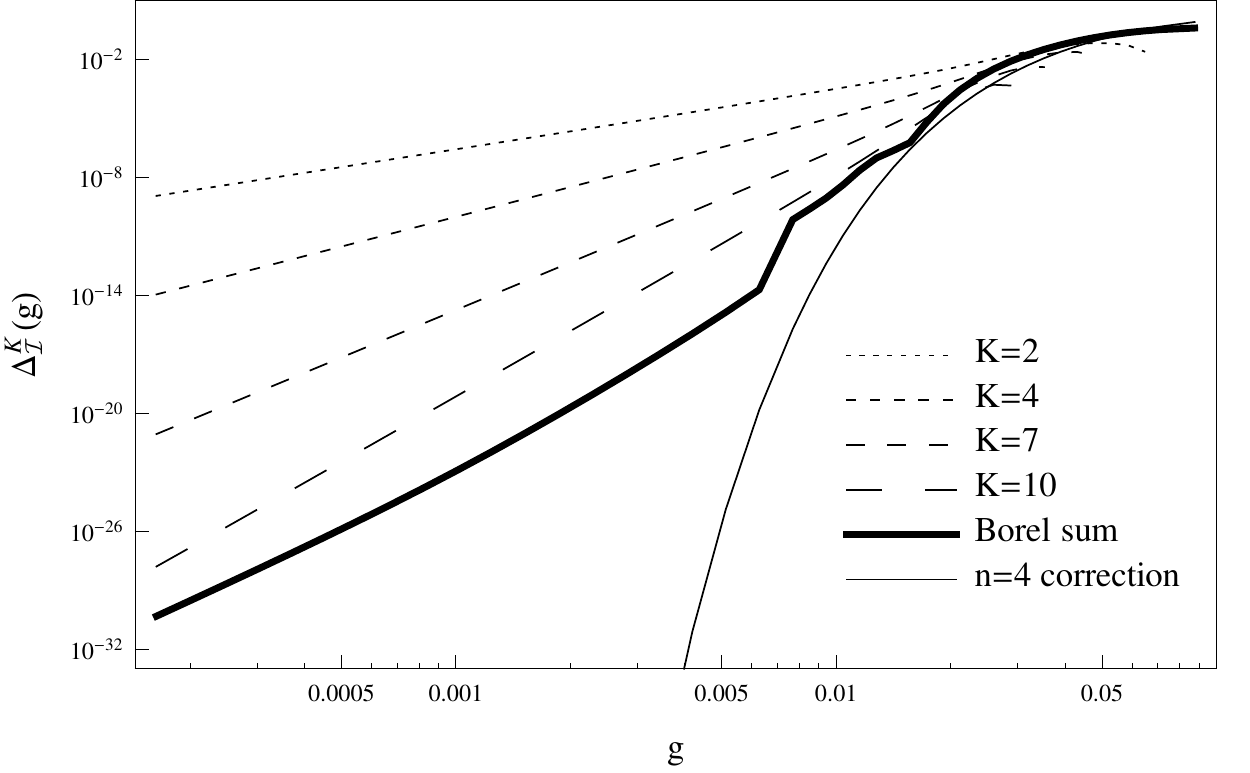}
\caption{Dashed lines represent function $\Delta_{\mathcal I}^K(g)$ for $K=2,4,7,10$. Imaginary ambiguity of energy decreases for growing $K$. Thick line is $\Delta_{\mathcal I}^{10}(g)$ with instanton contribution replaced by its Borel sum. Solid line is the leading term of the $n=4$ contribution in expansion (\ref{eq:full_expansion}) with $\epsilon_{430}=1$. Since it was neglected in our analysis, it is the lower bound for $\Delta_{\mathcal I}^{K}(g)$. It is approximately saturated for $g>0.02$.}\label{fig:imaginaryconvergence2}
\end{figure}

\begin{figure}
\centering
\includegraphics[width=\textwidth]{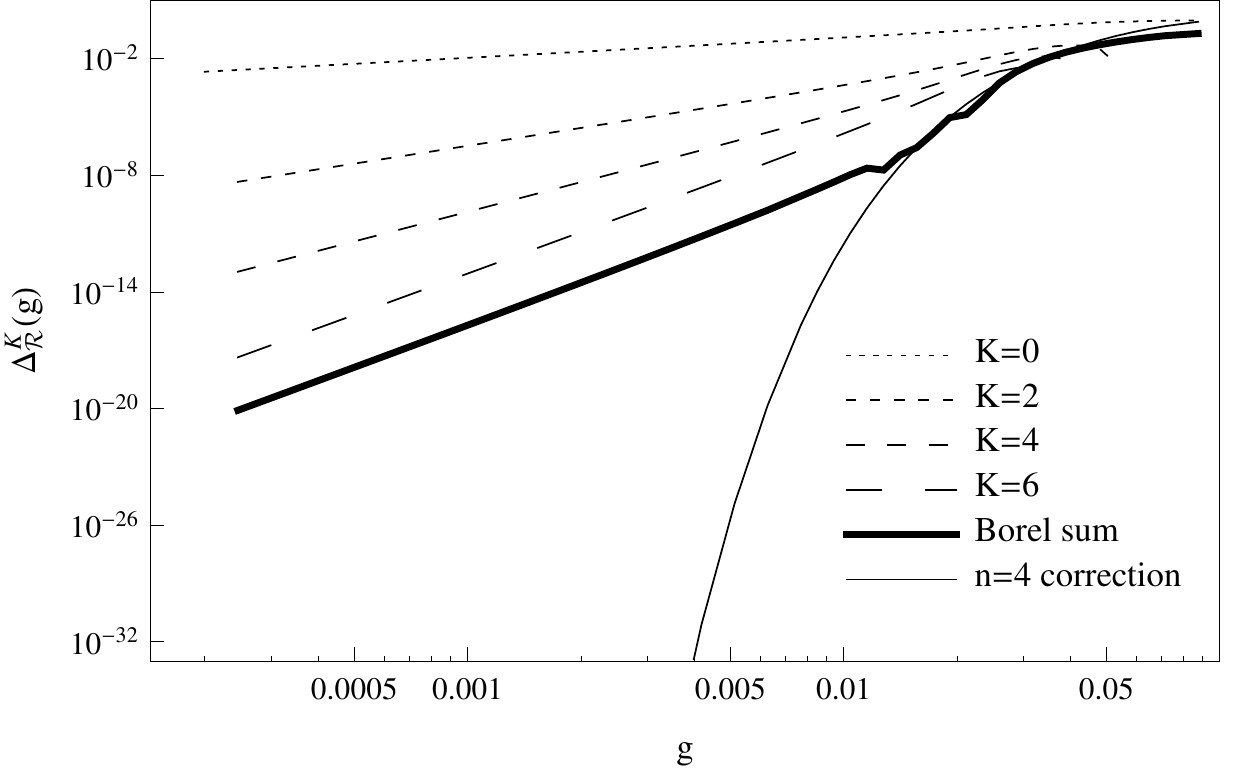}
\caption{Dashed lines represent function $\Delta_{\mathcal R}^K(g)$ for $K=0,2,4,6$. Approximation of energy improves for growing $K$. Thick line is $\Delta_{\mathcal R}^{6}(g)$ with instanton contribution replaced by its Borel sum. Thin line is $n=4$ contribution which is neglected in our analysis.}\label{fig:realconvergence3}
\end{figure}

In Fig. \ref{fig:imaginaryconvergence2} the plot of $\Delta_{\mathcal I}^K(g)$ with found coefficients is presented. One can observe that $\Delta_{\mathcal I}^K(g)$ decreases with growing $K$ and the relation $\Delta_{\mathcal I}^K(g)=\mathcal O(g^{K+1})$ holds. Having already a few coefficients of the $n=2$ expansion we found its Borel sum. It turns out that this procedure improves convergence by at least a factor of $10^2$ for $g\in(0.00016,0.009)$. From formula (\ref{eq:full_expansion}) it can be seen that $n=2$ terms do not have to remove full ambiguity of the energy. Some of it may be removed by $n=4$ contribution. Therefore, we cannot expect $\Delta_{\mathcal I}^K(g)$ to be smaller than $\frac{1}{g}e^{-1/3g}\ln^2(\frac{2}{g})$. This limitation is indeed seen in Fig. \ref{fig:imaginaryconvergence2}. Similar analysis was made for $\Delta_{\mathcal I}^K(g)$ and results are shown in Fig. \ref{fig:realconvergence3}. A detailed analysis concerning cancelation of higher order ambiguities was presented in \cite{UNSAL} for the cosine potential.


\section{Summary}
Using essentially exact numerical solutions of the anharmonic oscillator problem, we have verified the long-existing theoretical predictions that the ambiguities in resumming the perturbative, asymptotic series are nonperturbative, and indeed cancel with the ones from two-instanton interactions. The remaining, well defined part, is constructed additively from resummed perturbative series and nonperturbative contributions and is in agreement with our numerical results for $g<0.05$. For agreement in higher orders the story repeats itself on the level of multi-instanton interactions.

Thanks to high precision of computations we were able to confirm values of coefficients of two--instanton correction given by Zinn-Justin and Jentschura. It was also possible to estimate a few more coefficients of the energy expansion which appear in imaginary part of energy. Computing energies for smaller values of the coupling constant $g$ would require yet higher precision. For $g=10^{-5}$ the instanton correction is of order $10^{-14471}$ so precision of computations would have to be 16 times higher than for $g=0.00016$.

The series $E^{(2)}(g)$ appears to be asymptotic and not Borel summable. Considerable improvement of results is observed when on finds Borel sum of $E^{(2)}(g)$ even with only few coefficients of given expansion. Having more terms it might be possible to extract $E^{(4)}(g)$ contribution to the ground energy.

\subsection*{Acknowledgements}

We thank Mithat \"Unsal, Michael Teper and Yannick Meurice for fruitful discussions. This work was supported by Foundation for Polish Science MPD Programme co-financed by the European Regional Development Fund, agreement no. MPD/2009/6.
%\bibliography{bibliography}
%\input{bibliography.bbl}
%\begin{comment}
%\end{comment}

\end{document}